\input psfig

\documentstyle[12pt]{article}
\newcommand{\nc}{\newcommand}
\font\zfont = cmss10 %scaled \magstep1
\nc\ZZ{\hbox{\zfont Z\kern-.4emZ}}
\nc\bigone{\hbox{1\kern-.2eml}}
\nc{\ron}{\romannumeral}
\nc{\Bib}[1]{\bibitem{#1}}
\nc{\beq}{\begin{equation}}
\nc{\eeq}{\end{equation}}
\nc{\beqr}{\begin{eqnarray}}
\nc{\eeqr}{\end{eqnarray}}
\nc{\ket}[1]{| {#1}\rangle}
\nc{\bra}[1]{\langle {#1} |}
\nc{\psb}{\bar{\psi}}
\nc{\om}{\omega}
\nc{\omb}{\bar{\om}}
\nc{\theto}{\theta_{\circ}}
\nc{\bn}{\theta_{n}}
\nc{\thl}{\theta_{l}}
\nc{\zb}{\bar{z}}
\nc{\aj}{a_{j}}
\nc{\ej}{e_{j}}
\nc{\ejb}{\bar{e}_{j}}
\nc{\eplam}{\varepsilon_{\lambda}}
\nc{\adag}{a^{\dagger}}
\nc{\bdag}{b^{\dagger}}
\nc{\del}{\partial}
\nc{\delz}{\del_{z}}
\nc{\delzb}{\del_{\zb}}
\nc{\Cc}{\cal C}
\nc{\I}[2]{\int_{#1}d{#2}\,}
\nc{\minI}{\I{\Cc_{-}}{\theta}}
\nc{\pluI}{\I{\Cc_{+}}{\theta}}
\nc{\oI}{\I{\Cc_{\circ}}{\theta}}
\nc{\realI}{\int_{-\infty}^{\infty}d\theta\,}
\nc{\byp}{\frac{1}{2\pi}}
\nc{\bypi}{\frac{1}{2\pi i}}
\nc{\half}{\frac{1}{2}}
\nc{\Zpm}{Z_{\pm}}
\nc{\EL}{E_{L}}
\nc{\ER}{E_{R}}
\nc{\YL}{Y_{L}}
\nc{\YR}{Y_{R}}
\nc{\XR}{X_{R}}
\nc{\Bh}{B_{h}}
\nc{\Bmh}{B_{-h}}
\nc{\Fpm}{F_{\pm}}
\nc{\X}{\frac{X'}{1+X}}

\begin{document}
\rightline{RU-95-54}
\rightline{hep-th/9509071}
\vspace{0.4in}
\begin{center}
{\bf\Large Exact Partition Function and Boundary State of 2-D Massive Ising\\
Field Theory with Boundary Magnetic Field}
\end{center}
\vspace{0.6in}
\begin{center}
R. Chatterjee \footnote[1]{Email: robin@physics.rutgers.edu}\\
Department of Physics and Astronomy\\
Rutgers University\\
P.O.Box 849, Piscataway, NJ 08855-0849\\
\vspace*{0.4in}
{\Large\bf Abstract}
\end{center}

We compute the exact partition function, the universal ground state
degeneracy
and boundary state of the 2-D Ising model with boundary magnetic field at
off-critical temperatures. The model has a domain that exhibits states
localized near the boundaries. We study this domain of boundary bound
state and derive exact expressions for the ``$g$ function'' and boundary
state for all temperatures and boundary magnetic fields.
In the massless limit we recover the boundary renormalization group
flow between the conformally invariant free and fixed boundary conditions.

\vspace{0.4in}

\section{Introduction}
\vspace{0.1in}

Quantum field theory (QFT) in the presence of boundaries plays an
important role in our understanding of a wide range of physical phenomena.
Examples include various surface critical behaviors \cite{Binder}, impurity
effects (e.g. the Kondo problem) \cite{AL1,AL2}, junctions in quantum wires
\cite{Callan1}, dissipative quantum mechanics \cite{Callan2,Cald}, etc.
It is also
important in open string theory \cite{Love,W} where the open string
states are created
by vertex operators built from boundary operators in the boundary
conformal field theory.

An important concept in boundary QFT is the boundary state. In
Langrangian formulation the boundary condition may be
expressed either in terms of a boundary action or as a relevant
perturbation of a conformal boundary condition. On the space-time of a
cylinder of length $L$ and circle length $R$ with coordinate $x$
running along the length and coordinate $y$ running along the circle
(Fig.1), the action may be written as:
%in the sense of the former as:
\beq
{\cal A}= \int_{\cal D}d x d y\ {\cal L}_{\em bulk} + \int_{{\cal B}_l}
d y\ \Phi_{B_l}(y) + \int_{{\cal B}_r} d y\ \Phi_{B_r}(y)
\eeq
where $\Phi_{B_l}$ and $\Phi_{B_r}$ are boundary fields on the left and
right boundaries respectively. In Hamiltonian formulation there are two
alternative approaches: (1) In the first one chooses $x$ to be the space
coordinate
and $y$ the Euclidean time coordinate. The Hilbert space of states ${\cal
H}_{L}^{B_{l}B_{r}}$ associated with a $y = constant$ time slice must
satisfy boundary conditions $B_l$ and $B_r$ at the left and right
boundaries respectively.
For example, in the $L \to \infty$ limit the states in ${\cal
H}_{L}^{B_{l}B_{r}}$ may be classified as asymptotic scattering states.
Particles scatter with each other and off the boundaries. If the
scattering is factorizable (admitting boundary Yang Baxter symmetry
\cite{GZ}) then, relative to any boundary (and again in the limit of
large $L$), one may choose either a basis of asymptotic ``in'' states or of
asymptotic ``out'' states to span ${\cal H}_{L}^{B_{l}B_{r}}$. The two bases
are related by the factorizable scattering matrix. And the momenta of
these asymptotic states are constrained by the boundary conditions in
this Hamiltonian picture. The partition function is expressed as:
\beq
Z = Trace_{{\cal H}_{L}^{B_{l}B_{r}}}(e^{-R H_L})
\label{partR}
\eeq
where $H_L$ is the Hamiltonian,
and an n-point correlation function of local operators ${\cal
O}_{1}(x_{1},y_{1})$, $\ldots$, ${\cal O}_{n}(x_{n},y_{n})$ is given by:
\beq
\frac{1}{Z}\, Trace_{{\cal H}_{L}^{B_{l}B_{r}}}({\cal T}_{y}\,({\cal
O}_{1}(x_{1},y_{1})\ldots {\cal O}_{n}(x_{n},y_{n})\, e^{-R H_{L}}))
\eeq
where ${\cal T}_{y}$ implies ``y--ordering''.
(2) In the other approach $x$ is considered to be the Euclidean time
coordinate and $y$ the space coordinate. The Hilbert space ${\cal H}_R$
coincides with that of the theory with periodic (or antiperiodic)
boundary conditions (i.e. the theory in finite space without boundary).
The boundary conditions now appear as the initial state
$\ket{B_l}$ at time $x = 0$ evolving to the final state $\ket{B_r}$ at
time $x = L$. These states are called boundary states and encode all
information about the respective boundary conditions. The partition
function is now expressed as:
\beq
Z = \bra{B_r}e^{-L H_R} \ket{B_l} \label{partL}
\eeq
where $H_R$ is the Hamiltonian in this formulation. An n-point function
is expressed as:
\beq
\frac{\bra{B_r}{\cal T}_{x}\,{\cal O}_{1}(x_{1},y_{1})\ldots {\cal
O}_{n}(x_{n},y_{n}) \ket{B_l}}{\bra{B_r} B_l\rangle}
\eeq
where ${\cal T}_{x}$ now implies ``x--ordering''.

The two dimensional Ising model exhibits a rich variety of phenomena
many of whose characteristics can be calculated exactly. It has often
formed the basis
of our theoretical understanding of phase transitions and critical
phenomena, the renormalization group and general quantum field theory.
The model on a  manifold with boundaries and with a boundary magnetic
field has several interesting features. The boundary model is
integrable and its exact boundary scattering matrix is known for all
temperatures \cite{GZ}. At critical temperature, i.e. in the massless
case, the model describes a renormalization group (RG) flow from the
free to the fixed conformal boundary conditions \cite{Cardy1,AL1} and
exhibits a
monotonically decreasing universal ``$g$ function'' \cite{AL1} akin to
Zamolodchikov's $c$ function \cite{Z1} in the bulk. This g function
and the full
boundary state were computed exactly in \cite{Me}. They have also been
computed in \cite{Lecl1} using the thermodynamic Bethe ansatz (TBA) approach
for boundary theories \cite{AlZ,Sal}. In general, the boundary TBA
approach yields only the ratios
of the g factors due to some difficulties mentioned in \cite{Sal}. In
this paper, we compute the exact partition function, ground state
degeneracy factor g and boundary state for the massive case using the
approach in \cite{Me}.

\section{2-D Ising Model with Boundary Magnetic Field}
The 2-D Ising model with boundary magnetic field has been a subject of
investigation for quite some time. In \cite{M1, M2} (see also
\cite{Fisher}) the model was studied
on the lattice and the boundary contribution to the free energy and the
boundary spin spin correlation function were computed (see also
\cite{M3}). The free (i.e. boundary spins are free) and fixed (i.e. all
boundary spins are fixed to $+1$ or $-1$) limits of the boundary
condition were
studied in the context of boundary conformal
Field theory and the boundary spin operator was identified in
\cite{Cardy1, Cardy2} .
At critical temperature, the boundary RG flow from the free down to the
fixed
conformal boundary conditions was identified in \cite{AL1}. The model was
studied
in the context of boundary integrable field theory in \cite{GZ}. In
\cite{CZ} the local magnetization was computed exactly in the massless
theory using operator product expansions and were found to be
hypergeometric functions. Differential equations for the local energy
and magnetization have been derived recently in \cite{Lecl2}. The exact
partition function and boundary state were computed for the massless
theory in \cite{Me,Lecl1}.

As is known, the continuum theory of the Ising model is described by
free Majorana fermions.
On a manifold ${\cal D}$ with boundaries ${\cal B}_{i},\ i=1\ldots n$,
the model is described by the action:
\beqr
{\cal A} = \byp\I{\cal D}{x}d y\;[\psi\,\delzb\,\psi +
\psb\,\delz\,\psb + i
m \psi\psb]  \nonumber\\[0.1cm]
+ \sum_{j=1}^{n}\left\{\ \I{{\cal
B}_{j}}{t}[-\frac{i}{4\pi}\psi\psb + \half a\dot{a}] +
ih\I{{\cal B}_{j}}{t}a(t)(\ej^{\half}\psi +
\ejb^{\half}\psb)(t)\right\}
\label{eq:action}
\eeqr
Here $z = x + i y$ and $\zb = x - i y$ are the usual complex coordinates
on the manifold $\cal D$; $z = Z_{j}(t)$ and $\zb = \bar{Z}_{j}(t)$, with
$t$ being a real parameter,
parametrize each boundary ${\cal B}_{j}$; $\ej(t)=\frac{d Z_{j}}{d t}$,
$\ejb(t)=\frac{d \bar{Z}_{j}}{d t}$ are components of the tangent vector
$(\ej\ ,\ \ejb)$ to the $j$th boundary with
$\ej(t)\ejb(t)=1$; $\psi(x,y)$,
$\psb(x,y)$ are the free Majorana fermion fields and $m$ is the mass;
$a(t)=$ is a fermionic boundary field (see \cite{GZ,CZ}), anticommuting
with $\psi$ and $\psb$, with the two point function:
\beq
\langle a(t)\ a(t')\rangle_{free} = \half sign(t-t')
\eeq
and $\dot{a} \equiv \frac{d a }{d t}\,$; $h$ is the appropriately
rescaled external boundary magnetic field with the dimension of
$[length]^{\half}$. The integrand in the last term in (\ref{eq:action}) is
the boundary spin operator $\sigma_{B}$ (see \cite{GZ,Cardy1,Cardy2}):
\beq
\sigma_B(t) = a(t)(\ej^{\half}\psi + \ejb^{\half}\psb)(t)
\eeq
The first two terms in (\ref{eq:action}) comprise the bulk action and the
fermion mass $m$ is related to the temperature by
\beq
m \sim (T_{c} - T)
\eeq
As is known, the field theories corresponding to the ordered ($T < T_c$) and
disordered ($T > T_c$) phases are related by the duality transformation
$\psi \to \psi\ ,\ \psb \to -\psb$ and are equivalent. We
will assume the field theory for the low temperature phase i.e. $m > 0$
throughout this
paper. {}From (\ref{eq:action}) $\psi$, $\psb$ satisfy:
\beq
(\frac{\del}{\del x} + i\frac{\del}{\del y})\psi + i m \psb = 0
\eeq
\beq
(\frac{\del}{\del x} - i\frac{\del}{\del y})\psb - i m \psi = 0
\eeq
in the bulk and
\beq
(\frac{d}{d t} + i\lambda)\psi_{j}(t) = (\frac{d}{d t} -
i\lambda)\psb_{j}(t) \label{eq:bc}
\eeq
at each boundary ${\cal B}_j$, where
\beq
\lambda = 4\pi h^2
\eeq
and
\beq
\psi_{j}(t)= \ej^{\half}(t)\psi\big(Z_{j}(t)\big)\ , \qquad \psb_{j}(t)=
\ejb^{\half}(t)\psb\big({\bar Z}_{j}(t)\big)
\eeq
For our cylindrical geometry of Fig.1 boundary conditions
(\ref{eq:bc}) take the form:
\beq
({d\over dy}+ i\lambda){\bar\omega}\psi(0, -y)= ({d\over dy}-
i\lambda)\omega{\bar\psi}(0, -y)\hbox{\qquad at the left boundary $x = 0$,
\ and} \label{eq:bcleft}
\eeq
\beq
({d\over dy}+ i\lambda)\omega\psi(L, y)= ({d\over dy}-
i\lambda){\bar\omega}{\bar\psi}(L, y) \hbox{\qquad at the right
boundary $x = L$} \label{eq:bcright}
\eeq
where $\omega=e^{i\pi\over4}$ and ${\bar\omega}=e^{-i\pi\over4}$.

In the Hamiltonian picture (which from now on we call the
``{\bf L} channel'') where $x$ plays the role of Euclidean time
and $y$ is the spatial coordinate, $\psi$, $\psb$ admit the following
decompositions in terms of plane waves in the Neveu Schwarz (NS) (where
$\psi$, $\psb$ are antiperiodic along $y$) and Ramond (R) (where
$\psi$, $\psb$ are periodic along $y$) sectors:
\beq
\psi(x,y) = \sum_{\bn}\left[a(\bn)\,\om\, e^{\frac{\bn}{2}}e^{-m(\cosh{\bn}x +
i\sinh{\bn}y)} + \adag(\bn)\,\omb\, e^{\frac{\bn}{2}}e^{m(\cosh{\bn}x +
i\sinh{\bn}y)}\right] \label{mode1.1}
\eeq
\beq
\psb(x,y) = \sum_{\bn}\left[a(\bn)\,\omb\,
e^{-\frac{\bn}{2}}e^{-m(\cosh{\bn}x + i\sinh{\bn}y)} + \adag(\bn)\,\om\,
e^{-\frac{\bn}{2}}e^{m(\cosh{\bn}x + i\sinh{\bn}y)}\right]
\label{mode1.2}
\eeq
where
\beq
\sinh{\bn} =
\frac{2\pi}{mR}(n + \half) \ ,\; n \in \ZZ \; \mbox{ in the NS sector}
\eeq
and
\beq
\sinh{\bn} =
\frac{2\pi}{mR}n \ ,\; n \in \ZZ \mbox{ in the R sector}
\eeq
In (\ref{mode1.1}), (\ref{mode1.2})
$a(\bn)$ and $\adag(\bn)$ are the annihilation and creation
operators for a Fermi particle moving along the circle with momentum
$m\sinh{\bn}$ and energy
$m\cosh{\bn}$, $\bn$ being its rapidity, and satisfying standard
anticommutation relations.
A boundary state $\ket{B_h}$ at $x=0$ satisfies
\beq
\left[ a(-\bn) - i\tanh{\frac{\bn}{2}}\,\left(\frac{Q - \cosh{\bn}}{Q +
\cosh{\bn}}\right)\adag(\bn)\right]\ket{B_h} = 0 \label{eq:bc1.3}
\eeq
where
\beq
Q = (\frac{\lambda}{m} - 1) = (\frac{4\pi h^{2}}{m} - 1) \label{def:Q}
\eeq
and a similar equation holds for a boundary state at $x=L$. These are
obtained by imposing (\ref{eq:bcleft}) , (\ref{eq:bcright}) on
(\ref{mode1.1}) and (\ref{mode1.2}). Notice that the factor in the
second term in (\ref{eq:bc1.3}) is precisely the boundary scattering
matrix in the cross channel ${\cal K}(\bn) = {\cal
R}(\frac{i\pi}{2}-\bn)$ discussed in \cite{GZ}.

In the cross channel (henceforth we call it the ``{\bf R} channel'') in
which $y$
is the Euclidean time and $x$ is the
space coordinate, the mode expansions for $\psi$, $\psb$ are:
\beq
\psi(x,y)= \sum_{\thl}\left[ i\,
b(\thl)e^{-\frac{\thl}{2}}e^{m(i\sinh{\thl}x + \cosh{\thl}y)} +
\bdag(\thl)e^{-\frac{\thl}{2}}e^{-m(i\sinh{\thl}x + \cosh{\thl}y)}\right]
\label{mode2.1}
\eeq
\beq
\psb(x,y)= \sum_{\thl}\left[ -i\,
b(\thl)e^{\frac{\thl}{2}}e^{m(i\sinh{\thl}x + \cosh{\thl}y)} +
\bdag(\thl)e^{\frac{\thl}{2}}e^{-m(i\sinh{\thl}x + \cosh{\thl}y)}\right]
\label{mode2.2}
\eeq
for $y \in [0,R)$ and
\beq
\psi(x,R) = -\psi(x,0)\quad \mbox{and} \quad \psb(x,R) = -\psb(x,0)\quad
\mbox{in the NS sector},
\eeq
\beq
\psi(x,R) = \psi(x,0)\quad \mbox{and} \quad \psb(x,R) = \psb(x,0)\quad
\mbox{in the R sector}
\eeq
Here $b(\thl)$, $\bdag(\thl)$ destroy and create fermions (in the {\bf R}
channel) which now
move along $x$ with rapidity $\thl$. These fermions scatter off the
boundaries with the scattering matrix ${\cal R}(\theta)$ given by \cite{GZ}:
\beq
\bdag(-\theta) = {\cal R}(\theta)\,\bdag(\theta)\,;
\eeq
\beq
{\cal R}(\theta) = i\,\tanh{(\frac{i\pi}{4}
- \frac{\theta}{2})}\,\frac{Q + i\,\sinh{\theta}}{Q - i\,\sinh{\theta}}
\label{eq:Rmat}
\eeq
which can be obtained from (\ref{eq:bcleft}), (\ref{mode2.1}) and
(\ref{mode2.2}). The rapidities $\thl$ in (\ref{mode2.1}),
(\ref{mode2.2}) are solutions of:
\beq
1 + X = 0\;, \qquad \mbox{ where }\; X= e^{2 i m L\sinh{\theta}}
\tanh^{2}{(\frac{i\pi}{4}-\frac{\theta}{2})}\left(\frac{Q -
i\sinh{\theta}}{Q + i\sinh{\theta}}\right)^{\!2} \label{eq:basic}
\eeq
which we get from (\ref{eq:bcright}), (\ref{mode2.1}), (\ref{mode2.2})
and (\ref{eq:Rmat}). We notice that although $\thl = 0$ is a solution of
(\ref{eq:basic}) it is unphysical since ${\cal R}(0) = -1$ implying that
the wave function for this zero mode vanishes. In the next section we
start with the partition
function (\ref{partR}) as a trace over asymptotic states
(\ref{mode2.1}), (\ref{mode2.2}), (\ref{eq:basic}) and make explicit
transformations so as
to express it as a sum over states in the cross channel. In the large $L$
limit this will yield us the boundary state.

\section{Partition Function}
As mentioned in the earlier section, the spinor fields $\psi$, $\psb$
can be either periodic or antiperiodic along the circle of the cylinder.
We have been calling these the Ramond (R) and the Neveu Schwarz (NS) sectors
respectively. In the {\bf R} channel the states associated with the $y=0$
and $y=R$ lines are identical in the Ramond case while for the Neveu
Schwarz case they
differ by a phase of $\pi$. The partition function (\ref{partR}) in this
channel generalizes to $Z_{+}$ and $Z_{-}$ in the Neveu Schwarz and
Ramond sectors
respectively, where:
\beq
\Zpm = Trace_{{\cal{H}}_{L}^{B_{l}B_{r}}}\,((\pm 1)^{\cal F}\, e^{-R
H_{L}}) \,, \label{partiR}
\eeq
${\cal F}$ being the fermion number operator.
For our boundary Ising model (\ref{eq:action})
the partition function (\ref{partiR}) is given by:
\beq
\Zpm = e^{-R \EL}\prod_{\thl > 0}(1 \pm e^{- m R \cosh{\thl}})
\eeq
where $\EL$ is the ground state energy and $\theta_l\,$'s are solutions of
(\ref{eq:basic}). The free energy
$\Fpm$ is given by:
\beq
-R\Fpm = - R\EL + \sum_{\thl > 0} \log(1 \pm e^{- m R \cosh{\thl}})
\eeq
which can be written as (as in \cite{Me}):
\beq
- R\Fpm = -R\EL + \bypi \I{\cal C}{\theta}\frac{X'}{1+X}\log(1 \pm e^{- m
R \cosh{\theta}}) \label{eq:F1}
\eeq
where $X' = \frac{d X}{d \theta}$ and the contour ${\cal C}$ in the
complex $\theta$ plane is shown in
Fig.2 corresponding to the case when $Q > 1$. We note here that
$X(-\theta) = 1/X(\theta)$ and $X(i\pi - \theta) = X(\theta)$
The positions of zeroes and poles of $1 + X$ in the complex $\theta$
plane fall into three distinct domains: \\
({\ron 1}) $Q > 1$ : In this case all zeroes of $1 + X$ lie on the real
axis and
they cluster closer together further from the origin and with increasing
$L$. There are no poles in the upper half plane in the physical strip.
The picture is shown in Fig.2. \\
({\ron 2}) $0 < Q < 1$ : In this domain also all zeroes are on the real
axis as in ({\ron 1}). There is a pole on the negative imaginary axis at
$-i \sin^{-1}{Q}$ in addition to the ubiquitous pole at $-i\pi/2$
but none in the physical strip of the upper half plane. The analytic
structure is shown in Fig.3. \\
({\ron 3}) $ Q < 0$ : This domain is perhaps more interesting. There is
a pole on the positive imaginary axis at $\theta = - i \sin^{-1}Q$ and
possible zeroes on the imaginary axis as well. The pole corresponds to a
boundary bound state in the {\bf R} channel. We will discuss this domain in
greater detail later in the section. \\
Let us first consider domains ({\ron 1}) and ({\ron 2}) where all
solutions of (\ref{eq:basic}) are real and $1 + X$ has no pole in the
physical strip.

\subsection{\bf $Q > 0$ : Domain of No Boundary Bound State}
Here we treat domains ({\ron 1}) and ({\ron 2}) together. First we
compute $\EL$ :
\beq
\EL= -\half\sum_{\thl > 0}m\cosh{\thl}\quad =
-\frac{1}{4\pi i}\I{\cal C}{\theta}\frac{X'}{1+X}m\cosh{\theta}
\label{ELbegin}
\eeq
We can rewrite the integral on the right above as:
\beq
\EL= -\frac{1}{8\pi i}\left( \pluI + \minI - \oI \right)
\frac{X'}{1+X}m\cosh{\theta} \label{eq:EL}
\eeq
where the contours $\Cc_{+}$, $\Cc_{-}$ and $\Cc_{\circ}$ are shown in
Fig.4 and Fig.5. The last integral in (\ref{eq:EL}) is:
\beq
\frac{m}{8\pi i}\oI\X\cosh{\theta} =
\frac{m}{4} \label{eq:EL1}
\eeq
We rewrite the $\Cc_{-}$ integral in (\ref{eq:EL}) as:
\beq
\frac{m}{8\pi i}\minI\X\cosh{\theta} =
-\frac{m}{8\pi i}\minI\frac{X'\,\cosh{\theta}}{X^{2}(1+\frac{1}{X})} +
\frac{m}{8\pi i}\minI\frac{X'}{X}\cosh{\theta} \label{eq:EL0}
\eeq
After changing variable $\theta \to -\theta$ in the first term above and
expanding the second term we obtain:
\beq
\EL= \frac{m}{4} - \frac{1}{4\pi i}\pluI\X\cosh{\theta} -
\frac{m^{2}L}{2\pi}\int_{0}^{\infty}d\theta\cosh^{2}{\theta} -
\frac{m}{2\pi}\int_{0}^{\infty}d\theta\,[1 + \frac{2 Q
\cosh^{2}{\theta}}{Q^{2} + \sinh^{2}{\theta}}] \label{eq:EL2}
\eeq
The third term in (\ref{eq:EL2}) contains the nonuniversal bulk
intensive free energy which we set to zero. We recognize the fourth term
to be the boundary
intensive free energy and denote it by $\eplam$ :
\beq
2\eplam = -\frac{m}{2\pi}\int_{0}^{\infty}d\theta\,[1 +
\frac{2 Q\cosh^{2}{\theta}}{Q^{2} + \sinh^{2}{\theta}}] \label{eps}
\eeq
We now consider the second term in (\ref{eq:EL2}). We shift the contour
up by $\frac{i \pi}{2}$ (see Fig.6) (note that we can do this trivially
since the integrand has no singularity in this region), change variable
$\theta \to (\frac{i\pi}{2} - \theta)$ and integrate by parts to
obtain:
\beq
\frac{1}{4\pi i}\pluI\X\cosh{\theta} =
\frac{1}{4\pi}\int_{-\infty}^{\infty}d\theta
\cosh{\theta}\log(1 + Y_{L}(\theta)) \label{eq:EL3}
\eeq
where
\beq
Y_{L}(\theta) = X(\frac{i\pi}{2} - \theta)\quad = e^{-2 m L
\cosh{\theta}}\tanh^{2}{\frac{\theta}{2}}\left(\frac{Q -
\cosh{\theta}}{Q + \cosh{\theta}}\right)^{\!2} \label{YL}
\eeq
Finally we combine (\ref{eq:EL2}), (\ref{eps}), (\ref{eq:EL3}) to obtain:
\beq
\EL= \frac{m}{4} + 2\varepsilon_{\lambda} -
\frac{m}{4\pi}\int_{-\infty}^{\infty}d\theta\cosh{\theta}\log(1 +
\YL(\theta)) \label{ELfin}
\eeq

Now let us compute the second term in (\ref{eq:F1}). As we did for $\EL$, we
rewrite the integral as:
\beq
-R\Fpm= -R\EL + \frac{1}{4\pi i}\left(\pluI + \minI - \oI\right)
\X\log(1 \pm Y_R) \label{Fbegin}
\eeq
where
\beq
Y_{R} = e^{-m R \cosh{\theta}}
\eeq
The last integral is trivial:
\beq
\frac{1}{4\pi i}\oI\X\log(1 \pm \YR) = \half\log(1 \pm e^{-m R})
\label{F0}
\eeq
Once again, in the integral along contour $\Cc_{-}$, we rewrite in the
integrand:
\beq
\frac{X'}{1 + X} = -\frac{X'}{X^{2}(1 + \frac{1}{X})} + \frac{X'}{X}
\eeq
Then we transform variable
$\theta \to -\theta$ as we did in (\ref{eq:EL2}) and get:
\beqr
\frac{1}{4\pi i}\minI\X\log(1 \pm \YR) = \frac{1}{4\pi i}\pluI\X\log(1
\pm \YR) \nonumber\\[0.1cm]
+ \byp\int_{-\infty}^{\infty}d\theta\, m L \cosh{\theta} \log(1
\pm \YR) + \byp\int_{-\infty}^{\infty}d\theta\,
\left[\frac{1}{\cosh{\theta}} + \frac{2Q\cosh{\theta}}{Q^{2} +
\sinh^{2}{\theta}} \right]\log(1 \pm \YR) \label{F2}
\eeqr
We recognize the second term in (\ref{F2}) to be the Casimir energy:
\beq
E_{R}^{(\pm)} = -\frac{1}{2\pi}\realI m\cosh{\theta} \log(1 \pm
e^{-mR\cosh{\theta}}) \label{F3}
\eeq
where the $\pm$ signs, as usual, correspond to the NS and R sectors.
Next, for the first term in (\ref{F3}), we apply a Wick rotation of the
momentum axis in the complex plane by shifting the contour $\Cc_{+}$ in
the $\theta$ plane up by $i \pi/2$ (see Fig.6).
This enables us to go to the cross channel. Then we change variable
$\theta \to (\theta - \frac{i\pi}{2})$ and get:
\beq
\bypi\pluI \frac{X'}{1 + X} \log(1 \pm Y_{R}) = -\bypi\I{\Cc^{'}}{\theta}
\frac{\YL'}{1 + \YL} \log(1 \pm \XR) \label{eq:A}
\eeq
where the contour $\Cc^{'}$ is shown in Fig.7, $\YL$ is given by
(\ref{YL}) and
\beq
\XR(\theta) = \YR(\frac{i\pi}{2} - \theta) = e^{i m R \sinh{\theta}}
\eeq
After integrating by parts (\ref{eq:A}) yields:
\beq
\bypi\pluI\X \log(1 \pm \YR) = \bypi\I{\Cc^{'}}{\theta} \frac{\XR'}{\XR \pm
1} \log(1 + \YL)
\label{F4}
\eeq
We rewrite the integral along $\Cc{'}$
as a sum of integrals along $\Cc_{+}^{'}$ and $\Cc_{-}^{'}$ (see Fig.7)
\beq
\bypi\I{\Cc^{'}}{\theta}\frac{\XR'}{\XR \pm 1} \log(1 + \YL) =
\bypi\left(\I{\Cc_{+}^{'}}{\theta} + \I{\Cc_{-}^{'}}{\theta}\right)
\frac{\XR'}{\XR \pm 1} \log(1 + \YL) \label{Q}
\eeq
Our aim is to express the free energy as
a sum over states in the cross (i.e. {\bf L}) channel. So with
hindsight we change
variables $\theta \to -\theta$ in the $\Cc_{-}^{'}$ integral in (\ref{Q})
and obtain:
\beqr
\bypi\I{\Cc_{-}^{'}}{\theta}\frac{\XR'}{\XR \pm 1} \log(1 + \YL) =
\bypi\I{-\Cc_{-}^{'}}{\theta}\frac{\XR'}{\XR \pm 1} \log(1 + \YL)
\nonumber \\[0.1cm]
- \frac{m R}{2\pi}\int_{0}^{\infty}d\,\theta\cosh{\theta}\log(1 + \YL)
\eeqr
Thus (\ref{F4}) becomes:
\beqr
\bypi\pluI\X\log(1 \pm \YR) = \bypi\I{\Cc^{''}}{\theta}\frac{\XR'}{\XR
\pm 1} \log(1 + \YL) \nonumber \\[0.1cm]
- \frac{m R}{2\pi}\int_{0}^{\infty}d\,\theta\cosh{\theta}\log(1 + \YL)
\label{F5}
\eeqr
where $\Cc^{''} = \Cc_{+}^{'} + -\Cc_{-}^{'}$ is shown in Fig.8. The
first term above evaluates to:
\beqr
\bypi\I{\Cc^{''}}{\theta}\frac{\XR'}{\XR \pm 1} \log(1 +
\YL)\makebox[3.7in]{\hspace*{\fill}}
\nonumber\\[0.1cm]
\hspace{0.2in}= \left\{\begin{array}{c}\begin{displaystyle}
\log{\Sigma_{+}} =  \sum_{l \ge 0}\; \log(1 + \YL(\om_{l}))\;,\quad
m\sinh{\om_{l}} = \frac{2\pi}{R}l\;,\quad l \in \ZZ + \half\quad
\mbox{for}\; (+) \end{displaystyle}\\
\begin{displaystyle}\log{\Sigma_{-}} = \sum_{n \ge 0}\; \log(1 +
\YL(\om_{n}))\;,\quad
m\sinh{\om_{n}} = \frac{2\pi}{R}n\;,\quad n \in \ZZ\quad \mbox{for}\; (-)
\end{displaystyle}\end{array} \right\} \label{F6}
\eeqr
Finally combining (\ref{ELfin}), (\ref{Fbegin}), (\ref{F0}), (\ref{F2}),
(\ref{F3}), (\ref{F5}) and (\ref{F6}) we obtain:
\beqr
-R\Fpm = -L \ER^{(\pm)} - 2\varepsilon_{\lambda}R + \log{\Sigma_{\pm}} -
\half\log(e^{\frac{m R}{2}} \pm e^{-\frac{m R}{2}})  \nonumber \\[0.1cm]
+ \byp\realI
\left[\frac{1}{\cosh{\theta}} + \frac{2 Q \cosh{\theta}}{Q^{2} +
\sinh^{2}{\theta}}\right]\log(1 \pm e^{-m R \cosh{\theta}}) \label{Ffin}
\eeqr
We identify the last two terms as the universal ground state degeneracy:
\beqr
\log{g_{\pm}(R)} =
\frac{1}{4\pi}\realI \left[\frac{1}{\cosh{\theta}} + \frac{2 Q
\cosh{\theta}}{Q^{2} + \sinh^{2}{\theta}}\right]\log(1 \pm e^{-m R
\cosh{\theta}})  \nonumber \\[0.1cm]
- \frac{1}{4}\log(e^{\frac{m R}{2}} \pm e^{-\frac{mR}{2}})\label{g}
\eeqr
Thus from (\ref{Ffin}) and (\ref{g}) the partition function is:
\beq
Z_{\pm} = e^{-L\ER^{(\pm)} - 2\varepsilon_{\lambda} R}\; \Sigma_{\pm}\,
g_{\pm}^{2}(R) \label{Z}
\eeq
Expressions (\ref{Ffin}) and (\ref{Z}) with (\ref{F6}) and (\ref{g}) are
exact. Notice that in the $L \to \infty$ limit $\Sigma_{\pm} \to 1$.

Let us now look closely at the $Q < 0$ domain.

\subsection{$Q < 0$ : Domain Containing Boundary Bound State}

In the $Q < 0$ domain, i.e. for sufficiently weak boundary magnetic
field, the boundary scattering matrix \cite{GZ} has a pole in the
physical strip at $\theta = -i\sin^{-1}{Q}$. Correspondingly $1 + X$ has
a double pole at this same $\theta$. A careful analysis shows that, for
sufficiently large distance $L$ between the boundaries, there are also
two zeroes pinching the pole from above and below and approaching it
along the imaginary axis exponentially fast with increasing $L$. There
are also two more zeroes on the negative imaginary axis located
symmetrically to the upper ones and approaching $\theta = i\sin^{-1}{Q}$
at the same rate. There is however no pole there. These zeroes
correspond to states (in the {\bf R} channel) localized near the boundaries.
In fact there are two such states because the particle can be bound near
the left or the right boundary and the zero above the pole corresponds
to a symmetric (S) superposition of states localized near the left and
right boundaries while the zero below corresponds to their antisymmetric
(A) superposition. Each
boundary may be looked upon as an infinitely heavy particle sitting
there with the light fermions scattering off
it. At weak boundary magnetic fields they form bound states -- the
fermion gets trapped by the heavy boundary particle. In the limit of
large $L$ the S and A superposed states are degenerate (each bound
fermion does not see the opposite boundary and their wave functions do
not overlap) while, for finite $L$, they split by an amount exponentially
small in $L$ due to the overlap between the left and right localized
states (the bound fermion can now tunnel through the finite barrier and
become bound to the other boundary particle). Moreover, when $L$
decreases one zero each from above and below the real axis, which are
closer to the
origin, eventually leave the imaginary axis and go to the real axis
joining their friends there. The picture for sufficiently large $L$ is
shown in Fig.8. As we will see our results will not depend on $L$.

So now we include the contributions from the extra solutions to
(\ref{eq:basic}) in the integrals in (\ref{ELbegin}) and (\ref{Fbegin}).
We have:
\beq
\EL = -\frac{1}{4\pi i}\left[\,\I{\Cc}{\theta} + \I{\sigma_{1}}{\theta}
+ \I{\sigma_{2}}{\theta}\right] \frac{X'}{1 + X} m\cosh{\theta}
\label{newEL0}
\eeq
and
\beq
-R\Fpm = -R\EL + \bypi\left[\,\I{\Cc}{\theta} + \I{\sigma_{1}}{\theta}
+ \I{\sigma_{2}}{\theta}\right] \frac{X'}{1 + X} \log(1 \pm e^{-m R
\cosh{\theta}}) \label{newF0}
\eeq
where $\sigma_{1}$ and $\sigma_{2}$ are small closed contours enclosing
the zeroes below and above the pole respectively in the upper half
plane. We repeat the same steps as before in computing the integrals along
contour $\Cc$. However, when we shift the integration contour $\Cc_{+}$
up by $\frac{i\pi}{2}$ in obtaining (\ref{eq:EL3}) and (\ref{eq:A}) we
have to
avoid the singularities of $\frac{X'}{1 + X}$ in the integrand at the
two zeroes and the (double) pole of $(1 + X)$ (see Fig.9). In the
process we pick up contributions from each of these singularities (see
Fig.10):
\beq
\bypi\pluI \frac{X'}{1 + X}(\cdots) = \bypi\left[
\I{\Cc_{+}^{W}}{\theta} - \I{\sigma_{1}}{\theta} -
\I{\sigma_{2}}{\theta} - \I{\sigma_{pole}}{\theta}\right] \frac{X'}{1 +
X}(\cdots) \label{D}
\eeq
Here $\Cc_{+}$ is the same as in (\ref{eq:EL}) and (\ref{Fbegin}). The last
two
integrals cancel exactly with corresponding integrals in (\ref{newEL0})
and (\ref{newF0}) and we are left with the integral along
$\sigma_{pole}$ only. Thus the extra contribution, in effect, comes only
from the pole, irrespective of the number of zeroes on the imaginary
axis (of course we have to include all these zeroes in (\ref{newEL0}),
to start with). In the light of the above discussion, our result is therefore
independent of $L$ as long as we include a $\sigma$ contour for each zero
(of $1 + X$) in the physical strip. Thus we have:
\beq
\frac{1}{4\pi i}\I{\sigma_{pole}}{\theta}\frac{X'}{1 + X} m\cosh{\theta}
= - m R \cos{u} \label{toEplam}
\eeq
where
\beq
u = - \sin^{-1}{Q} \;, \qquad Q < 0
\eeq
Expression (\ref{toEplam}) gives the bound state contribution to the
boundary energy. For $Q <
0$ the intensive boundary energy (let us now denote it by $\eplam^{b}$)
is therefore:
\beq
2\eplam^{b} = -\frac{m}{2\pi}\int_{0}^{\infty}d\theta\,[1 +
\frac{2 Q\cosh^{2}{\theta}}{Q^{2} + \sinh^{2}{\theta}}] - m R
\cos{u} \;,\qquad Q < 0 \label{newEplam}
\eeq
And we have:
\beq
-\bypi \I{\sigma_{pole}}{\theta}\frac{X'}{1 + X} \log(1 \pm e^{-m R
\cosh{\theta}}) = 2\log(1 \pm e^{-m R \cos{u}}) \label{to_g}
\eeq
implying that now in the presence of boundary bound state in the {\bf R}
channel, the {\bf L} channel ground
state degeneracy function (which we denote by $g_{\pm}^{b}$ for $Q < 0$) is:
\beqr
g_{\pm}^{b}(R) = \frac{1}{4\pi}\realI \left[\frac{1}{\cosh{\theta}} +
\frac{2 Q
\cosh{\theta}}{Q^{2} + \sinh^{2}{\theta}}\right]\log(1 \pm e^{-m R
\cosh{\theta}})  \nonumber \\[0.1cm]
- \frac{1}{4}\log(e^{\frac{m R}{2}} \pm e^{-\frac{mR}{2}}) + \log(1 \pm
e^{-m R \cos{u}}) \;, \qquad Q < 0\label{new_g}
\eeqr
In an alternative interpretation, we can consider $g_{\pm}^{b}$ as an
analytic continuation of $g_{\pm}$ given in (\ref{g}) with the integral
in (\ref{g}) now being along the contour ${\cal C^{(Q)}}$ shown in
Fig.11 instead of along
the real axis. The contour surrounds the two poles of the integrand at
$\pm i\,\sin^{-1}{Q}$. Notice that the term in $[\cdots]$ in the
integrand is precisely the
derivative of the phase shift from boundary scattering and appears also
in the boundary thermodynamic Bethe ansatz approach \cite{Sal}.
$g_{\pm}$ defined this way covers the whole domain of $Q$ and we treat
this as our general definition for $g_{\pm}$. Denoting this by
$g_{\pm}^{(Q)}$, we
have:
\beq
g_{\pm}^{(Q)}(R) = \frac{1}{4\pi}
\I{\Cc^{(Q)}}{\theta}\left[\frac{1}{\cosh{\theta}} + \frac{2 Q
\cosh{\theta}}{Q^{2} + \sinh^{2}{\theta}}\right]\log(1 \pm e^{-m R
\cosh{\theta}}) - \frac{1}{4}\log(e^{\frac{m R}{2}} \pm
e^{-\frac{mR}{2}}) \label{Def_gQ}
\eeq
\beqr
g_{\pm}^{(Q)}(R) =  \left\{\begin{array}{c}\begin{displaystyle}
g_{\pm}(R)\;, \qquad Q \ge
0 \end{displaystyle} \vspace{0.2cm}\\[0.1cm]
\begin{displaystyle} g_{\pm}^{b}(R) \;, \qquad Q < 0
\end{displaystyle}\end{array}\right.
\eeqr
where the contour $\Cc^{(Q)}$ is shown in Fig.11. In the same way, we
define the intensive boundary energy $\eplam^{(Q)}$
for all $Q$:
\beq
2\eplam^{(Q)} = -\frac{m}{4\pi}\I{\Cc^{(Q)}}{\theta} [1 +
\frac{2 Q\cosh^{2}{\theta}}{Q^{2} + \sinh^{2}{\theta}}] \label{Def_epQ}
\eeq
\beqr
\eplam^{(Q)} = \left\{ \begin{array}{c}\begin{displaystyle} \eplam \;,
\qquad Q \ge 0
\end{displaystyle} \vspace{0.2cm}\\[0.1cm]
\begin{displaystyle} \eplam^{b} \;, \qquad Q < 0
\end{displaystyle}\end{array} \right.
\eeqr
The partition function in the whole domain of $Q$ is therefore given by:
\beq
Z_{\pm}^{(Q)} = e^{-L\ER^{(\pm)} - 2 R\,\eplam^{(Q)}}\,\Sigma_{\pm}\,
(g_{\pm}^{(Q)}(R))^2  \label{Def_ZQ}
\eeq

We will now compute the boundary state in the next section.

\section{Boundary State}
We consider the low temperature phase $T < T_c$ of the Ising system as
we have been doing. As we discussed earlier the boundary state belongs
to the {\bf L} channel Hilbert space ${\cal H}_{R}$. In this space there are
two vacuum states $\ket{0}$ and $\ket{\sigma}$ lying in the
Neveu-Schwarz and Ramond sectors respectively. In the infinite volume
(large $R$) limit the allowed momenta are continuous. The ground states
$\ket{0}$ and $\ket{\sigma}$ become degenerate. They are respectively
the symmetric and antisymmetric combinations of the vacuum states
$\ket{0_+}$ and $\ket{0_-}$ which are the broken $Z_2$ symmetry ground
states and correspond to all spins pointing up and all spins pointing
down respectively:
\beq
\ket{0} = \frac{1}{\sqrt{2}}\,(\,\ket{0_+} + \ket{0_-}\,)\, , \qquad
\ket{\sigma} = \frac{1}{\sqrt{2}}\,(\,\ket{0_+} - \ket{0_-}\,)
\label{vac}
\eeq
If $\Pi$ is the overall spin flip operator, we have
$\Pi\,\ket{0} = \ket{0}$ and $\Pi\,\ket{\sigma} = - \ket{\sigma}$.
Obviously $Z_+$ and $Z_-$ represent contributions from the $\Pi = 1$ and
$\Pi = -1$ respectively. We expect that:
\beq
Z_{\pm}^{(Q)} = Z_{hh} \,\pm\, Z_{h-h}
\eeq
where
\beq
Z_{hh} = \bra{B_{h}}\,e^{-L H_R}\, \ket{B_{h}} \,\quad \mbox{and} \quad
Z_{h-h}= \bra{B_{h}}\,e^{-L H_R}\, \ket{B_{-h}}
\eeq
implying that
\beq
Z_{h h} = \half(\, Z_{+}^{(Q)} + Z_{-}^{(Q)}\,) \quad \mbox{and} \quad
Z_{h-h} = \half(\,Z_{+}^{(Q)} - Z_{-}^{(Q)}\,) \label{Zh1}
\eeq
For large $L$ we have:
\beq
Z_{hh} = e^{-L \ER^{(+)}}(\bra{0}B_{h}\rangle)^2 + e^{-L
\ER^{(-)}}(\bra{\sigma}B_{h}\rangle)^2 \label{Zh2}
\eeq
\beq
Z_{h-h} = e^{-L \ER^{(+)}}(\bra{0}B_{h}\rangle)^2 - e^{-L
\ER^{(-)}}(\bra{\sigma}\Bh\rangle)^2 \label{Zh3}
\eeq
where we have used $\bra{0}\Bmh\rangle = \bra{0}\Bh\rangle$ and
$\bra{\sigma}\Bmh\rangle = -\bra{\sigma}\Bh\rangle$ and we have chosen
the phase of $\ket{0}$ to be such that $\bra{0}\Bh\rangle$ and
$\bra{\sigma}\Bh\rangle$ are real (see \cite{AL1}). Thus, from
(\ref{Zh1}) -- (\ref{Zh3}) it follows that:
\beq
\bra{0}\Bh\rangle = \frac{1}{\sqrt{2}}\, e^{-R\,\eplam^{(Q)}}
g_{+}^{(Q)}(R) \quad
\mbox{and} \quad \bra{\sigma}\Bh\rangle = \frac{1}{\sqrt{2}}\,
e^{-R\,\eplam^{(Q)}} g_{-}^{(Q)}(R) \label{Bh1}
\eeq
As was shown in \cite{GZ}, the presence of
an infinite set of
integrals of motion for a system with boundary makes the boundary
state take a simple form with contributions from pairs of particles of
opposite momenta (and possibly a zero mode) only.
Following \cite{GZ} and as was done in \cite{Me} for the critical case,
we write down the boundary state from (\ref{mode1.1}), (\ref{mode1.2})
and (\ref{Bh1}):
\beqr
B_{\pm h} = \frac{1}{\sqrt{2}} e^{-R\,\eplam^{(Q)}}\left[ g_{+}^{(Q)}(R)
\exp\left\{\sum_{n=0}^{\infty} {\cal K}(\theta_{n +
\half})\,\adag(\theta_{n +
\half})\,\adag(-\theta_{n + \half})\right\}\,\ket{0}\right.
\nonumber\\[0.1cm]
\pm \left.g_{-}^{(Q)}(R)\, \exp\left\{\sum_{n=1}^{\infty} {\cal
K}(\theta_{n})\,\adag(\theta_{n})
\,\adag(-\theta_{n})\right\}\,\ket{\sigma}\right]
\label{state}
\eeqr
where $\sinh{\theta_{j}} = \frac{2\pi}{m R}j$, $g_{\pm}^{(Q)}(R)$ and
$\eplam^{(Q)}$ are
given by (\ref{Def_gQ}) and (\ref{Def_epQ}) respectively, and from
(\ref{eq:Rmat}),
\beq
{\cal K}(\theta) = {\cal R}(\frac{i\pi}{2} - \theta) \quad =\quad
i\tanh{\frac{\theta}{2}}\,\left(\frac{Q - \cosh{\theta}}{Q +
\cosh{\theta}}\right)
\label{K}
\eeq

\section{Asymptotic Limits}

\underline{\large{$m R \to 0$}} \\[0.1cm]
We can take the (bulk) ultraviolet limit $mR \to 0$ in two ways: ({\ron
1}) we can either make $m \to 0$ i.e. increase the temperature $T \to
T_{c}$ on the same cylinder, or ({\ron 2}) we can reduce the size $R$ of
the system $R \to 0$ while maintaining the same temperature so that the
system size becomes much smaller compared to its correlation length.

Let us discuss case ({\ron 1}) first. As $m \to 0$, $Q$
increases, the boundary bound state, if initially present, disappears
and $g_{\pm}^{(Q)} = g_{\pm}$. The theory now describes a renormalization
group flow in the space of boundary interactions interpolating between
the free and the fixed conformal boundary conditions \cite{AL1,Cardy1}.
The massless particles group into right
and left movers which do not see each other. We have to obtain the $mR
\to 0$ limits of ${\cal R}(\theta)$ and ${\cal K}(\theta)$ carefully. We
do this by shifting the rapidity occurring in ${\cal R}(\theta)$ by infinite
amounts: $\theta \to \log{\frac{2}{mR}} + \theta$  for the right movers and
$\theta \to -\log{\frac{2}{mR}} - \theta$ for the left movers. For the
continuation of $\cal R$ in the cross channel, i.e. ${\cal K}(\theta)$,
however, we must shift $\theta \to - \log{\frac{2}{mR}} - \theta$ for
right movers and $\theta \to \log{\frac{2}{mR}} + \theta$ for the left
movers instead to obtain the correct ${\cal K}(k) = {\cal K}(i k)$, $k$
being the momentum of the massless fermions (e.g. for right movers $k =
\frac{1}{R}e^{\theta}$).
Moreover, in the $mR \to 0$ limit the integral in (\ref{g}) falls into
three regions: $({\romannumeral 1})\; \theto \ll
\log{\frac{2}{mR}}$, i.e. $h^{2}R \ll 1$ where $\theto =
\sinh^{-1}\sqrt{Q^{2} - 2}$ is an extremum of $\Phi(\theta) =
\frac{2Q\cosh{\theta}}{q^{2} + \sinh^{2}{\theta}}\,$, $({\romannumeral
2})\; \theto \sim \log{\frac{2}{mR}}$, i.e. $h^{2}R \sim 1$ and
$({\romannumeral 3})\; \theto \gg \log{\frac{2}{mR}}$, i.e.
$h^{2}R \gg 1$
It is easy to see from (\ref{g}) that in region $({\romannumeral 2})$
the following hold:
\beq
g_{+}(R) = \frac{\sqrt{2\pi}}{\Gamma(\alpha +
\half)}\left(\frac{\alpha}{e}\right)^{\alpha}
\eeq
\beq
g_{-}(R) = \frac{2^{1/4} \sqrt{\pi \alpha}}{\Gamma(\alpha +1)}\left(
\frac{\alpha}{e}\right)^{\alpha}
\label{g:m=0}
\eeq
\beq
\mbox{where}\qquad \alpha = 2 h^{2} R
\eeq
with regions $({\romannumeral 1})$ and
$({\romannumeral 3})$ above appearing as $h^{2}R \to 0$ and $h^{2}R \to
\infty$ limits of (\ref{g:m=0}) respectively:
\beq
g_{+} = \sqrt{2} \quad \mbox{ and }
\quad g_{-} = 0 \qquad \mbox{ for }\; h^{2}R \to 0 \;, \label{g:a=0}
\eeq
\beq
g_{+} = 1 \quad \quad \mbox{ and } \quad g_{-} = 2^{-\frac{1}{4}} \quad
\mbox{ for }\; h^{2}R \to \infty \label{g:a=inf}
\eeq
We then obtain the boundary state in the massless limit from
(\ref{state}), (\ref{g:m=0}) and (\ref{K}):
\beqr
\ket{B_{\pm h}} = e^{-R\,\eplam}\left(\frac{\alpha}{e}\right)^{\alpha}
\sqrt{\pi}\left[\frac{1}{\Gamma(\alpha + \half)}\exp\left\{
i\,\sum_{n=0}^{\infty} \frac{n + \half - \alpha}{n + \half + \alpha}
\adag_{n+\half}\bar{a}^{\dagger}_{n+\half} \right\}\,\ket{0} \right.
\nonumber \\[0.1cm]
\pm \left.\frac{2^{1/4}\sqrt{\alpha}}{\Gamma(\alpha + 1)} \exp\left\{
i\,\sum_{n=1}^{\infty} \frac{n - \alpha}{n + \alpha} \adag_{n}
\bar{a}^{\dagger}_{n} \right\}\,\ket{\sigma}\right] \label{state:m=0}
\eeqr
where $\adag_{j}$, $\bar{a}^{\dagger}_{j}$ are creation operators for right
and left movers of momentum $k_{j} = \frac{2\pi}{R}j$. Expressions
(\ref{g:m=0}) and (\ref{state:m=0}) were
obtained directly
in \cite{Me, Lecl1} in the case of the critical Ising model. As was shown in
\cite{Me}, at
the conformal limits, (\ref{state:m=0}) yields for:\\
\underline{$h^{2}R \to 0$ :}
\beq
\ket{B_{free}} = \exp(i\,\sum_{n=1}^{\infty} \adag_{n+\half}
\bar{a}^{\dagger}_{n+\half})( \ket{0} - \ket{\varepsilon}) \label{state:free}
\eeq
where $\ket{\epsilon}$ is the energy density operator in the bulk
conformal field theory and we used:
\beq
\adag_{\half}\,\bar{a}^{\dagger}_{\half}\,\ket{0} = i\,\ket{\varepsilon}
\eeq
and for \\
\underline{$h^{2}R \to \infty$ :}
\beqr
\ket{B_{fixed\pm}} = \frac{1}{\sqrt{2}} \exp(-i\,\sum_{n=1}^{\infty}
\adag_{n+\half} \bar{a}^{\dagger}_{n+\half}) (\ket{0} + \ket{\varepsilon})
\nonumber \\[0.1cm]
\pm \;\frac{1}{2^{1/4}} \exp(-i\,\sum_{n=1}^{\infty} \adag_{n}
\bar{a}^{\dagger}_{n}) \ket{\sigma} \label{state:fixed}
\eeqr
Both (\ref{state:free}) and (\ref{state:fixed}) agree with Cardy's
results \cite{Cardy1}

Now we consider the case ({\ron 2}) way of taking the $mR \to 0$ limit,
namely we let $R \to 0$ while keeping the temperature constant.
For finite $Q$ it is easy to see that as $R \to 0$ the system reaches
the conformally invariant limit of free boundary conditions. We readily
see that in this limit
(\ref{Def_gQ}) yields $g_{+} = \sqrt{2}$ and $g_{-} = 0$ for both $Q \ge
0$ and $Q < 0$ domains (here we note that for $Q < 0$ the second term in
the integral in (\ref{new_g}) yields $\log{\frac{1}{\sqrt{2}}}$ instead
of $\log{\sqrt{2}}$ that one obtains for $Q \ge 0$) and expression
(\ref{state:free}) for the boundary state.

\vspace{0.3cm}
\noindent\underline{\large{$m R \to \infty$}}\\[0.1cm]
Now we consider the case $mR \to \infty$ when the size of the
system is much larger compared to its correlation length. {}From
(\ref{Def_gQ}) we readily see that in this limit,
\beq
g_{+}^{(Q)}(R) = g_{-}^{(Q)}(R) \sim e^{-\frac{mR}{8}}
\eeq
{}From (\ref{Def_gQ}) and (\ref{state}) it is easy to obtain :
\beq\ket{B_{\pm h}} = e^{-\frac{mR}{8}}\,\frac{1}{\sqrt{2}}(\ket{0} \pm
\ket{\sigma}) + excited\; states
\eeq
as we expect since these are just the broken symmetry vacua $\ket{0_+}$
and $\ket{0_-}$ (\ref{vac}) discussed in the previous section.

\section{Conclusion}
Expressions (\ref{Def_gQ}) for the ``g function'' and (\ref{state}) for
the boundary state are
the most important results of this paper. We have computed these
quantities by starting with
the partition function expressed as a sum over states (satisfying the
boundary conditions) in one channel and making explicit transformations
so as to express it as a sum over states in the cross channel. The domain
$Q < 0$, in which there are states localized near the boundaries (in the
{\bf R} channel), is particularly interesting. We found single expressions
for the g function and the boundary state valid in all domains of $Q$ by
analytic continuation, whereby the contour of integration is made to
surround some relevant poles.
It would be interesting to see if this could form a general prescription
for treating boundary bound states in obtaining the $g$ function in
general integrable quantum field theory with boundary,
and it would be desirable to gain a proper physical understanding
underlying this treatment.

\vspace{0.4in}
\begin{center}
{\Large\bf Acknowledgement}
\end{center}
I am especially grateful to A. Zamolodchikov for numerous illuminating
discussions, his support and inspiration and valuable comments on the
manuscript. I also wish to thank V.\nolinebreak\ Brazhnikov, K.
Intrilligator and
M. Zyskin for valuable discussions.

\newpage
\begin{figure}[htbp]
\vspace*{-1.1in}
\centerline{\psfig{file=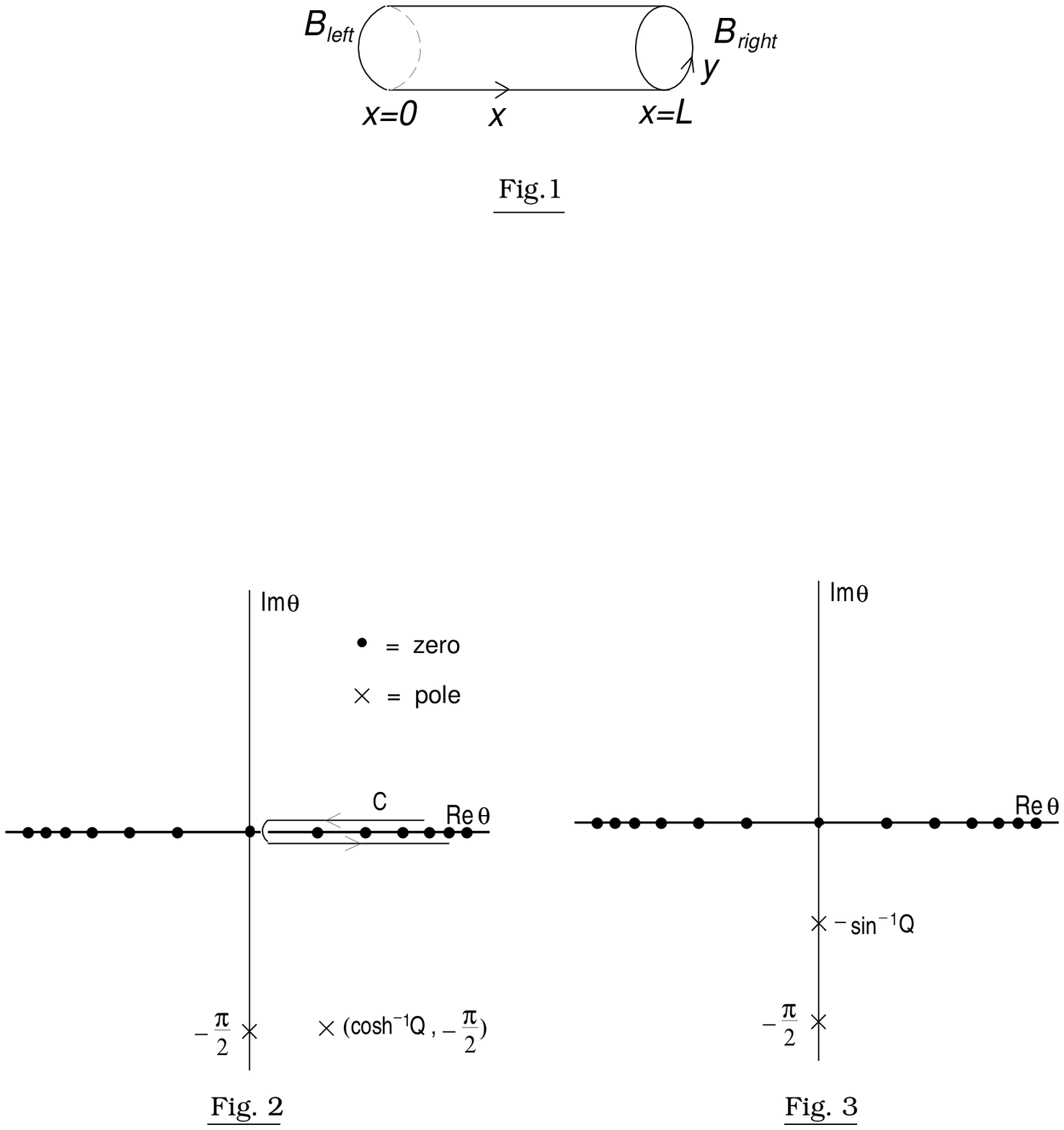}}
\end{figure}
\newpage
\begin{figure}[htbp]
\vspace*{-1.1in}
\centerline{\psfig{file=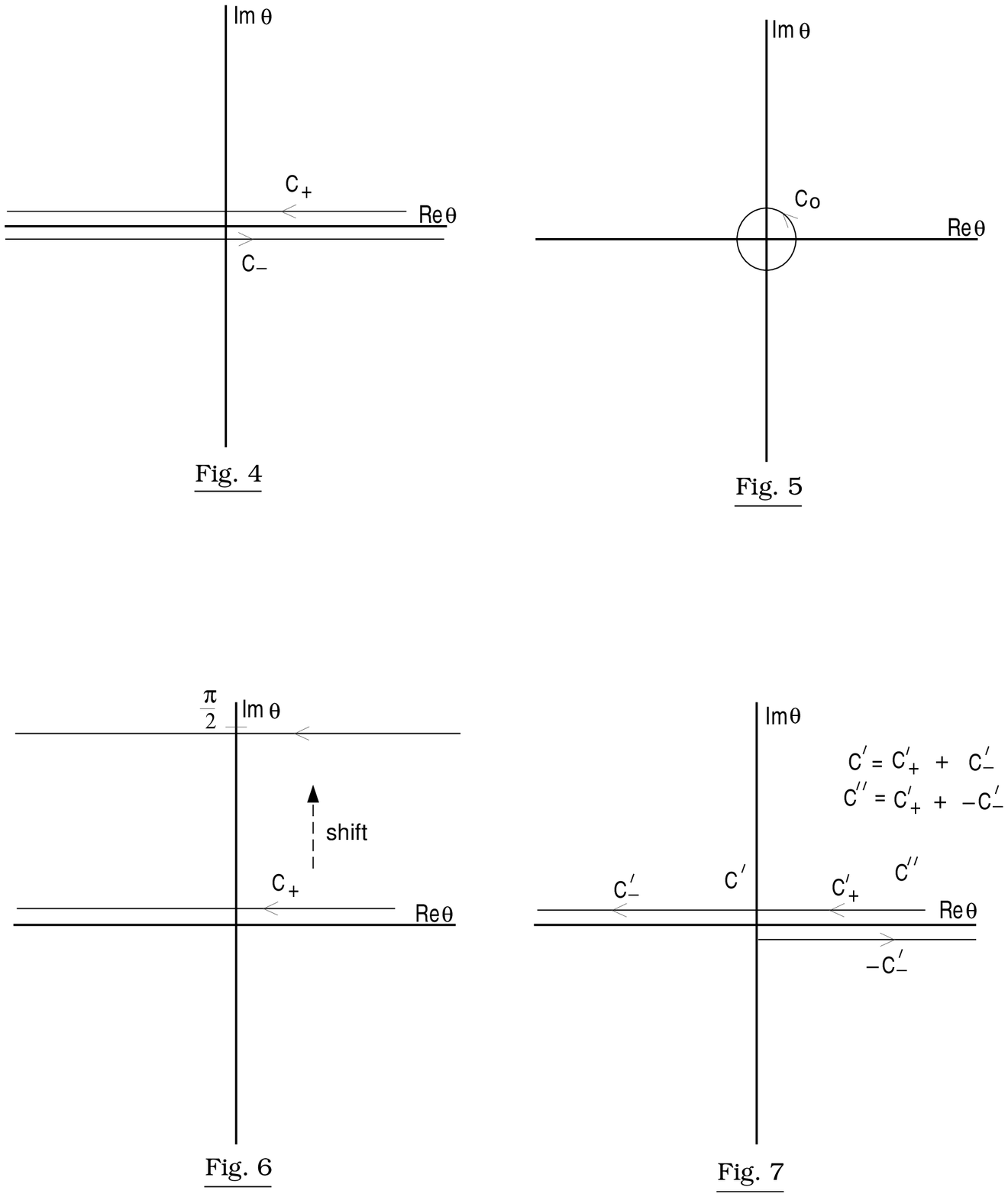}}
\end{figure}
\newpage
\begin{figure}[htbp]
\vspace*{-1.1in}
\centerline{\psfig{file=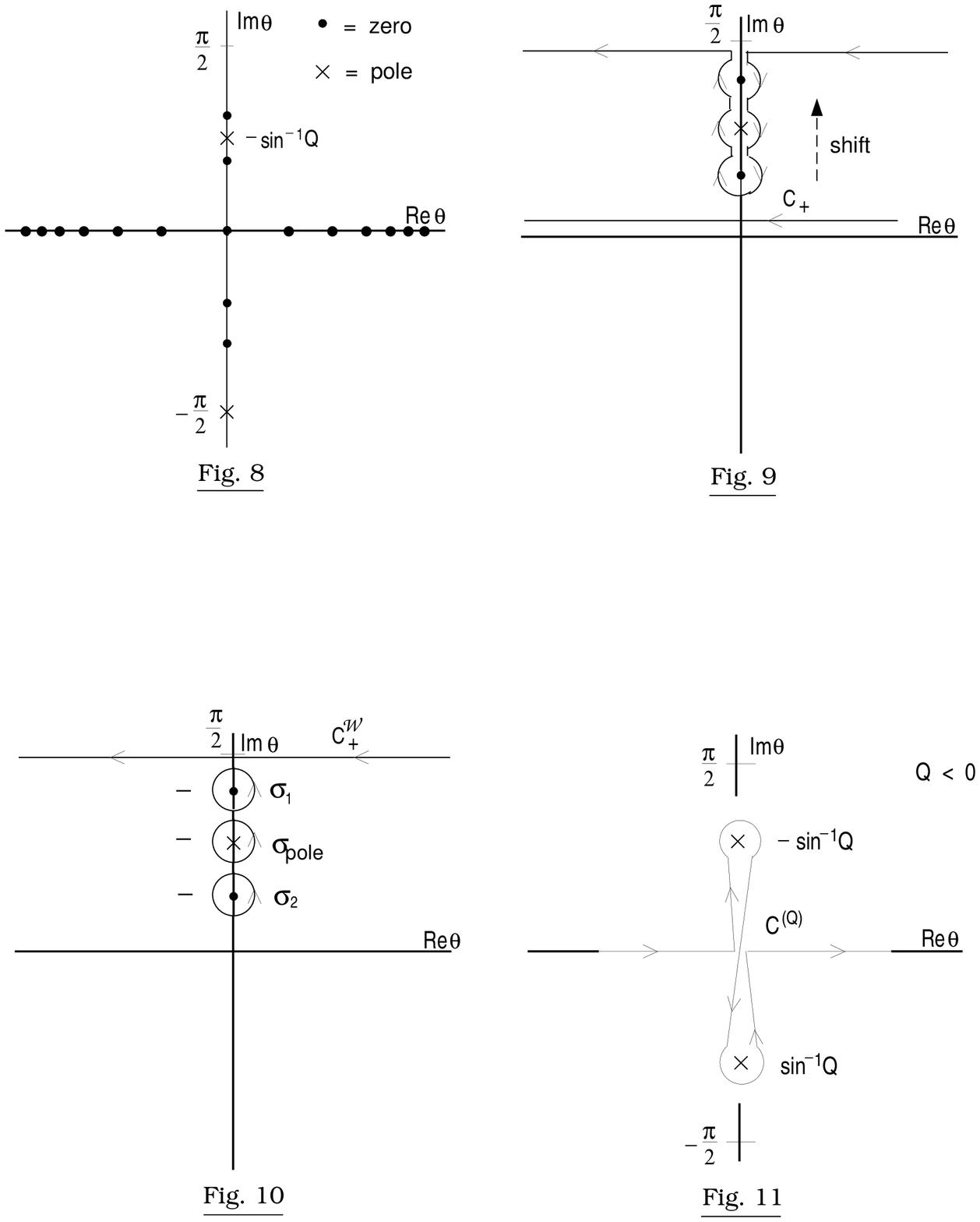}}
\end{figure}
\end{document}